\documentclass[12pt,preprint]{aastex}

\usepackage{graphicx}

\newcommand{\beq}{\begin{equation}}
\newcommand{\eeq}{\end{equation}}
\newcommand{\beqa}{\begin{eqnarray}}
\newcommand{\eeqa}{\end{eqnarray}}
\newcommand{\pp}{^{\prime\prime}}

\def\la{\lower.5ex\hbox{$\; \buildrel < \over \sim \;$}}
\def\ga{\lower.5ex\hbox{$\; \buildrel > \over \sim \;$}}

\begin{document}

\title{A Dynamical Model of the Local Group}

\author{P.~J.~E. Peebles}  
\affil{Joseph Henry Laboratories, Princeton University, Princeton, NJ 08544}
\author{R. Brent Tully}
\affil{Institute for Astronomy, University of Hawaii, 2680 Woodlawn Drive, Honolulu, HI 96822}
\author{Edward J. Shaya}
\affil{Department of Astronomy, University of Maryland, College Park, MD 20742}

\begin{abstract}
This dynamical model for the 28 galaxies with  distances less than 1.5 Mpc, and not apparently tight satellites, is constrained by the initial condition that peculiar velocities at high redshift are small and growing in accordance with the standard cosmology. The solution is a satisfactory fit to most of the measured redshifts, distances, and proper motions, with some interesting exceptions that call for further investigation. The model predicts Milky Way rotation speed 256 km~s$^{-1}$, consistent with Reid et al. (2009a). Ten Local Group galaxies emanate from low supergalactic latitude and supergalactic longitude $\sim 70^\circ$, perhaps as remnants from failed assembly of a larger galaxy. NGC 6822 passes close to the Milky Way at redshift $z\sim 0.27$, in an orbit similar to the Magellanic Clouds. Leo~I has heliocentric angular velocity $0.33$~mas~yr$^{-1}$, perhaps measurable by the mean stellar motion, and 15 galaxies have proper motions greater than $0.05$~mas~yr$^{-1}$, measurable for any with masers. 
\end{abstract}
\maketitle

\section{Introduction}\label{sec:1}

Advances in the measurements of distances and proper motions of nearby galaxies are producing a rich data set and the opportunity to explore dynamics in the local universe in much greater detail than was possible a decade ago (as in Peebles, Phelps, Shaya, \& Tully 2001). This is a report of progress in the analysis of what happened within the Local Group (LG) based on the redshift and distance data in the Local Universe Catalog, also a work in progress (RBT and EJS), and on the measured proper motions of three Local Group (LG) members. The dynamical model includes orbits for the Milky Way (MW) and the 27 galaxies with measured distances less than 1.5 Mpc, after excluding the Small Magellanic Cloud, whose orbit may be entangled with the Large Magellanic Cloud (LMC), and excluding the galaxies that are judged to be too close to MW or M31 for a feasible analysis of where they were at high redshift. The circular velocity $v_c$ of MW at the Solar circle also is a parameter, mainly for the conversion between heliocentric and galactocentric velocities (as noted by Shattow \& Loeb 2009). 

The  starting condition from cosmology is that the peculiar velocities of the matter at high redshift are generally smaller than now and increasing in accordance with the linear perturbation theory of growing departures from a homogeneous mass distribution of an ideal pressureless fluid. The resulting numerical solutions for the orbits of LMC and M33 at low redshift are similar to solutions that ignore this initial condition (Besla et al. 2007; Kallivayalil {\it et al.} 2009; Putman  et al. 2009), because the present positions and velocities  are similar. Orbits computed with and without the cosmological initial condition diverge at high redshift, however, in analogy to the difference at high redshift between a growing mode and a mixture of growing and decaying modes in perturbation theory. LG analogs derived in numerical simulations (as in Boylan-Kolchin,  Besla, \&  Hernquist  2010;  Libeskind et al. 2011) also satisfy the cosmological initial condition, and are an important complement to this explicit analysis of what is observed in the Local Group.

\section{The Model}\label{sec:methods}

\subsection{Approximations}

This analysis depends on three major approximations. The first is the representation of  each LG galaxy by a point particle back to the starting time of the computation, at expansion factor $1+z_i=10$. The physical picture is that each particle is meant to track the mean motion of the fragments that are gathering and merging into a protogalaxy, and to represent the gravitational attraction of all these fragments on neighboring protogalaxies. 

The second major approximation is the initial condition at $z_i$. In linear perturbation theory at high redshift, $z$, the growing density contrast in an ideal pressureless fluid scales as $\delta\rho/\rho\propto 1/(1+z)$, and the peculiar velocity $\vec v$ and acceleration $\vec g$ are related to the expansion time $t$ from $z\rightarrow\infty$ by 
\beq
\vec v=\vec gt.
\label{eq:gt}
\eeq
We apply this initial condition to the particle model. An equivalent condition is that the comoving coordinate displacements from the particle positions at $t=0$ scale as $\vec x(t)-\vec x(0)\propto a(t)$.  This is an approximation, among other reasons because it does not take account of the nonlinear development of structure on the scale of galaxies at $z_i$. The point is that it fits the idea that peculiar velocities of protogalaxies at high redshift are small and  growing as structure grows. 

The third approximation is the use of four external actors for a phenomenological description of the effect of matter external to LG on the motions of the LG galaxies. The actors represent in first approximation the neighboring mass concentrations in the Sculptor Group and the Maffei and Centaurus systems, but the masses, positions and redshifts of the actors are allowed to float to improve the model fit to the measurements of the LG galaxies. The approximation would fail if a large external mass that now has little effect on LG was a serious influence in the past, when these masses were much closer to LG. This need not be a problem, of course, because LG was more compact too. And we have a meaningful indication that this and the other two approximations yield a useful approach to reality, from the generally reasonable fit of the model to considerably more measurements than there are relevant adjustable parameters. But an important goal for a more complete analysis is to check the third approximation by taking explicit account of the dynamics and gravitational effects of the observed external mass concentrations. 

\subsection{Numerical Method}

In the form applied here the numerical action method (NAM) of dealing with the mixed  boundary conditions of given present positions and the initial condition in equation~(\ref{eq:gt}) represents particle orbits by positions at discrete time steps  (Peebles 1989, 1995; Peebles, Phelps, Shaya, \& Tully 2001). It produces a solution to the equation of motion in leapfrog approximation that is reached by iterated joint shifts of all particle positions from random trial orbits toward a stationary point of the action in the direction indicated by the first and second derivatives of the action. The method is described  in Appendix A. 

The mixed boundary conditions allow multiple solutions, and there is no guarantee that our adoption of  the best fit to the constraints chooses the most realistic solution. Thus the numerical solution presented here certainly may have some wrong orbits. Wrong orbits for the smaller galaxies that behave as test particles may be discovered from improved measurements of positions and velocities, or perhaps from signatures of disturbances by close passages of more massive galaxies. The orbits of more massive galaxies are more tightly  constrained by their gravitational effects on other orbits, but that is not a guarantee of uniqueness. Also to be born in mind is that our $\chi^2$ measure of fit to the constraints as a function of the parameters has many local minima. Here again there is no guarantee that we have arrived at the minimum corresponding to the most accurate solution. The straightforward way to explore both ambiguities --- multiple solutions and multiple minima of the measure of fit to the data --- is to repeat  construction of the numerical solution starting from many different random orbits. This was only very partially done for the solution presented here, in the following procedure. 

The construction commenced with dynamical solutions for eight galaxies: MW and M31, the three galaxies with measured proper motions --- LMC, M33, and IC10 --- and with NGC~3109, NGC~300, and the Maffei system serving in effect as external actors. The small particle number made it practical to find many solutions starting from random orbits (quadratic curves with random coefficients that allow the orbits to arc over several megaparsecs) and measured parameters randomly chosen within their assigned standard deviations. In the few solutions with promising fits to the measurements all the free parameters were adjusted to minimize a $\chi^2$ measure of fit. The solution that ended up with the smallest $\chi^2$ was the starting point for the addition of galaxies one at a time. Each addition started with a large set of trial random orbits for the new galaxy, and each trial orbit relaxed to a solution to the equation of motion for all the galaxies. In the solution with the smallest $\chi^2$ all parameters for all particles were adjusted to minimize $\chi^2$. In some cases this cycle was repeated many times in a search for the best fit to the measurements. 

The value of $\chi^2$ as a function of the masses, present positions, and $v_c$ has discontinuities where a slight change in a parameter causes the solution to jump to quite different orbits, generally with much larger $\chi^2$. To deal with this the solution was stored before each trial adjustment of a parameter, and the solution and parameter restored if the adjustment crossed a discontinuity indicated by a distinctively large number of iterations to a solution to the equation of motion.  A parameter shift that increased $\chi^2$, or encountered a discontinuity, was multiplied by 0.75 and the sign changed for the next round of parameter shifts. A shift that lowered $\chi^2$ was multiplied by 1.25 for the next iteration.

It would be informative to repeat this whole procedure with different random orbits and a different sequence of addition of the galaxies. The practical problem with the numerical method used here is that the computation time to add a galaxy to the solution scales as the fourth power of the number,  $n_p$, of galaxies (varying as $n_p^3$ for the joint relaxation to a solution and as $n_p$ to explore how the parameters of all the galaxies in the solution are to be adjusted to minimize the $\chi^2$ measure of fit). Also, with increasing $n_p$ the discontinuities in $\chi^2$ appear more frequently, and discovering them requires more iterations in the relaxation to a stationary point of the action. The result was computation time of about a week (on an iMac quad) to add the last two galaxies, Sextans~A and~B, in the solution presented here. Use of the present numerical method to explore the many local minima of $\chi^2$ and the many choices of multiple solutions certainly would not be undertaken lightly. But the numerical method is a work in progress, along with the measurements of distances and velocities. 

\subsection{Parameters}

Since the model is at best a useful approximation --- galaxies are not point particles --- it cannot be expected to yield a highly precise fit to the galaxy positions and velocities. The strategy for dealing with this along with the measurement uncertainties is to assign what seem to be moderately optimistic goals for the differences between model and measurements, use these goals as effective standard deviations in a $\chi^2$ measure of fit, and adjust parameters to minimize this $\chi^2$. 

\begin{table}[htpb]
\centering
\begin{tabular}{llrrcrcr}
\multicolumn{8}{c}{Table 1: The Local Group Galaxies}\\
\noalign{\medskip}
\tableline\tableline\noalign{\smallskip}
 & Name  & $cz_{\rm cat}$  &  ${\Delta cz^{\rm a}}$ & 
    $D_{\rm cat}$ & $\Delta D^{\rm a}$ &  $D_\perp$  & $v_i$  \\
   \noalign{\smallskip}
\tableline
\noalign{\smallskip}
  1&    MW&\nodata  &  \nodata &$ 0.0085 $& \nodata &  \nodata &  33\\
  2&   LMC&$  271$ & $ -1.1 $&$ 0.049\pm  0.01$&$  0.007$&  0.003&  87\\
  3& Leo I&$  229$ & $-11.4 $&$ 0.26\pm  0.01$&$ -0.019$&  0.004&  33\\
  4& Leo T&$   35$ & $  1.5 $&$ 0.42\pm  0.02$&$ -0.005$&  0.000& 101\\
  5&   Phx&$  -13$ & $ -7.8 $&$ 0.43\pm  0.02$&$  0.034$&  0.001&  40\\
  6& N6822&$  -57$ & $ -0.6 $&$ 0.51\pm  0.03$&$ -0.011$&  0.000&  40\\
  7&  N185&$ -227$ & $  9.8 $&$ 0.64\pm  0.03$&$ -0.135$&  0.001&  37\\
  8&  LGS3&$ -281$ & $ -0.3 $&$ 0.65\pm  0.13$&$  0.028$&  0.000&  51\\
  9&Cet dS&$  -87$ & $-34.2 $&$ 0.73\pm  0.04$&$  0.221$&  0.004&  36\\
 10&  N147&$ -193$ & $  0.5 $&$ 0.73\pm  0.04$&$  0.026$&  0.000&  39\\
 11& Leo A&$   28$ & $ -0.0 $&$ 0.74\pm  0.11$&$  0.010$&  0.000&  36\\
 12& I1613&$ -238$ & $  0.8 $&$ 0.75\pm  0.04$&$  0.005$&  0.000&  51\\
 13&  AXIV&$ -481$ & $  0.1 $&$ 0.78\pm  0.10$&$ -0.039$&  0.000&  38\\
 14&Cas dS&$ -307$ & $  0.2 $&$ 0.79\pm  0.04$&$  0.005$&  0.000&  35\\
 15&  IC10&$ -348$ & $ -6.3 $&$ 0.79\pm  0.04$&$  0.213$&  0.001&  59\\
 16&   M31&$ -297$ & $ 12.0 $&$ 0.79\pm  0.03$&$  0.052$&  0.000&  26\\
 17&  AXII&$ -556$ & $ -1.2 $&$ 0.83\pm  0.05$&$  0.139$&  0.000&  47\\
 18&   M33&$ -180$ & $-11.5 $&$ 0.92\pm  0.05$&$ -0.239$&  0.000&  53\\
 19&Tucana&$  194$ & $-37.1 $&$ 0.92\pm  0.05$&$  0.450$&  0.003&  17\\
 20&PegDIG&$ -178$ & $  0.1 $&$ 0.95\pm  0.05$&$ -0.079$&  0.000&  59\\
 21&DDO210&$ -132$ & $-29.7 $&$ 0.97\pm  0.06$&$  0.423$&  0.002&  27\\
 22&   WLM&$ -124$ & $  3.1 $&$ 0.98\pm  0.05$&$ -0.010$&  0.000&  33\\
 23&SagDIG&$  -73$ & $-29.1 $&$ 1.05\pm  0.05$&$  0.341$&  0.002&  22\\
 24& N3109&$  403$ & $ -9.6 $&$ 1.33\pm  0.07$&$  0.285$&  0.000&  49\\
 25&Antila&$  361$ & $  1.8 $&$ 1.35\pm  0.07$&$ -0.019$&  0.000&  50\\
 26& U4879&$  -27$ & $  7.7 $&$ 1.36\pm  0.03$&$ -0.034$&  0.000&  32\\
 27& Sex B&$  302$ & $ -7.3 $&$ 1.43\pm  0.07$&$  0.268$&  0.000&  39\\
 28& Sex A&$  325$ & $ -2.8 $&$ 1.43\pm  0.07$&$  0.138$&  0.000&  42\\
 \noalign{\smallskip}
\tableline
\noalign{\smallskip}
\multicolumn{8}{l}{$^{\rm a}$ model minus catalog value; \ \ units: km s$^{-1}$, Mpc} \\
\end{tabular}
\end{table}

The 28 LG galaxies listed in Table~1 are all the entries in the  Local Universe Catalog that are closer than 1.5~Mpc and far enough away from MW and M31 that their orbits are judged likely to be computable in NAM. The catalog angular positions agree with the NASA/IPAC Extragalactic Database (NED), the redshifts and distances occasionally are more recent than the NED entries, and many of the catalog distance uncertainties are quite different from the NED entries.

The effective standard deviations in the LG galaxy redshifts are taken to be 5 km s$^{-1}$ for M31, which plays a central dynamical role, and for LMC, whose proper motion is a key constraint, with the looser goal of 10 km s$^{-1}$ difference between model and measurement for all the other LG galaxies. This nominal uncertainty is far larger than the precision of some redshift measurements, but we judge it to to be optimistic for our numerical model. The assigned effective standard deviations in distances are  6~kpc for LMC, 50~kpc for M31, and twice the stated uncertainties in Table~1 for the difference between model and measured distances of the rest of the LG galaxies. Again, this is arguably optimistic considering the limitations of the model and the possibility that the difficult art of distance measurements has allowed some errors well above the  estimated uncertainties. Offsets between model and measured present positions perpendicular to the line of sight may be real --- a galaxy of stars may not be centered on its dark matter halo --- or perhaps more likely a fault of the approximate model. The assigned standard deviations in the perpendicular offsets are $D_\perp = 2$ kpc for LMC, because a larger error causes an objectionable error in angular position of this nearby galaxy,  $D_\perp = 1$ kpc for M31, which partially sets the model orientation, and $D_\perp = 5$ kpc for all the other LG galaxies.  The logarithms of the particle masses enter $\chi^2$ with standard deviations equivalent to a factor of ten. This very loose constraint on masses allows some indication of how well the model might predict the masses that are most relevant to the dynamics, and which of the galaxies have masses so small, as indicated by the luminosities in the Local Universe Catalog, that their masses do not matter. Each orthogonal component of each initial peculiar velocity at $1+z_i=10$ is allowed standard deviation 50~km~s$^{-1}$ in $\chi^2$. This choice is based on the consideration that since the protogalaxies were close to touching at $1+z_i=10$ the protogalaxy peculiar motions might be expected to have been comparable to motions within protogalaxies. The constraint on initial velocities is needed also to prevent NAM from producing an occasional solution with quite unreasonably large $v_i$. 

The conversion between  galactocentric and heliocentric velocities uses the fixed Solar velocity components $U=11.1$, $V=12.2$, $W=7.2$~km~s$^{-1}$ relative to the local standard of rest (Sch{\"o}nrich, Binney \& Dehnen  2010). The circular velocity of the local standard of rest is a  parameter to be adjusted. In the computation of $\chi^2$ we adopt the nominal or catalog value $v_c=230\pm 10$ km s$^{-1}$, intermediate between a standard estimate, $220$~km~s$^{-1}$, and the larger result obtained by Reid et al. (2009a). As discussed in the next section the minimization of $\chi^2$ favors a circular velocity larger than nominal and close to Reid et al. Coordinate positions and peculiar velocities are referred to the Friedmann-Lema\^\i tre cosmology with flat space sections, Hubble parameter  $H_o=70$~km~s$^{-1}$~Mpc$^{-1}$,  matter density parameter  $\Omega_m=0.27$, and radiation ignored. 

The value of $v_c$ also figures in the gravitational acceleration produced by MW, which is modeled as $g= v_c^2/r$ at distance $r<r_o= G m/v_c^2$, where $m$ is the MW mass, with the inverse square law at larger separation. This is relevant for LMC and NGC\,6822. The mass distribution in M31 is modeled the same way, with circular velocity fixed at $250$~km~s$^{-1}$, but as it happens no model orbits pass within $r_o$ for M31. All other galaxies are treated  as point particles. None pass close enough to any galaxy except MW to suggest a serious problem with this.

\begin{table}[htpb]
\centering
\begin{tabular}{lrrrrr}
\multicolumn{6}{c}{Table 2: Proper Motions}\\
\noalign{\medskip}
\tableline\tableline\noalign{\smallskip}
  & \multicolumn{2}{c}{$\mu_\alpha$} & \  &  \multicolumn{2}{c}{$\mu_\delta$}  \\
   \noalign{\smallskip}
  \cline{2 - 3}  \cline{5 - 6}
 & measured & model  &\ & measured & model  \\
\tableline
 \noalign{\smallskip}
 LMC$^{\rm a}$ & $2.03 \pm 0.08$ & 2.10 &\  & $0.44 \pm 0.05$ & 0.45 \\
 M33$^{\rm b}$ & $23.0 \pm  6.0$  & 16.1 &\ & $2.0 \pm 7$ & 6.4 \\ 
 IC10$^{\rm b}$ & $-2.0 \pm 8$ & 6.7 &\  & $20.0 \pm 8.0$ & $-3.5$\\
\tableline
 \noalign{\smallskip}
\multicolumn{6}{l}{$^{\rm a}$milli arc sec y$^{-1}$\quad $^{\rm b}$micro arc sec y$^{-1}$}\\
\end{tabular}
\end{table}

We have constraints on the proper motions from Kallivayalil {\it et al.} (2006) and Piatek, Pryor \& Olszewski (2008) for LMC,  Brunthaler et al. (2005) for M33, and  Brunthaler et al. (2007) for IC10. Table~2 lists the adopted proper motions as the components $\mu_\alpha$ and  $\mu_\delta$ of the heliocentric angular velocity in the directions of increasing right ascension $\alpha$ and increasing declination $\delta$. The adopted angular velocity of LMC is from Kallivayalil {\it et al.} (2006). The heliocentric angular velocities of M33 and IC10 are based on measured motions of masers that must be corrected for the motions of the masers relative to the host galaxies, and the conversion from velocities within the galaxy to angular velocities depends on the distance to the galaxy. Distances are adjustable parameters in this analysis, but since the effect of changing the distance within a reasonable range is small we adopt the correction at distance 809~kpc for M33, with a straight mean of the proper motions from the two masers, and distance 760~kpc for the one maser in IC10. The measurement uncertainties are derived from the stated uncertainties in the angular velocities of the masers and in their motions within the galaxies. We treat the uncertainties in Table~2 as standard deviations in the computation of $\chi^2$.

\begin{table}[htpb]
\centering
\begin{tabular}{llrrcrcrr}
\multicolumn{9}{c}{Table 3: External Actors}\\
\noalign{\medskip}
\tableline\tableline\noalign{\smallskip}
  & Name & $cz_{\rm cat}$  &  $\Delta cz^{\rm a}$ & 
    $D_{\rm cat}$ & $\Delta D^{\rm a}$ &  $D_\perp$  & $v_i$ & 
   $\log(m)$ \\
   \noalign{\smallskip}
\tableline
\noalign{\smallskip}
 29&  N300&$  141$ & $ -7.8 $&$ 2.08\pm  0.10$&$  0.035$&  0.020&  32&  11.38\\
 30&   N55&$  129$ & $-17.0 $&$ 2.11\pm  0.10$&$  0.088$&  0.037&   9&  11.91\\
 31&Maffei&$   22$ & $ 22.7 $&$ 3.20\pm  0.15$&$ -0.128$&  1.071&   5&  12.56\\
 32&   Cen&$  547$ & $  7.5 $&$ 3.57\pm  0.15$&$ -0.225$&  0.686&  20&  12.01\\
 \noalign{\smallskip}
\tableline
\noalign{\smallskip}
\multicolumn{9}{l}{$^{\rm a}$model minus catalog value; \ \ units: km s$^{-1}$, Mpc, Solar masses} \\
\end{tabular}
\end{table}

The external actors in Table~3 are nominally the observed mass concentrations  in the nearby Maffei and Centaurus systems, with the galaxies NGC~55 and NGC~300 assigned to represent the still closer Sculptor group. It should be understood that numerical values of parameters assigned to these external actors are not to be taken as useful approximations to the properties of the real systems: they are meant to provide a phenomenological description of the effect of external mass on the LG. For this purpose we allow greater latitudes in parameters. Each is assigned effective standard deviation of 50 km s$^{-1}$ in redshift and a factor of 50 in mass. NGC 55 and NGC 300 are assigned effective standard deviations of 200 kpc in radial distance and 100 kpc in perpendicular offset, and the actors representing the Maffei and Centaurus systems are assigned standard deviations of 300 kpc in distance and 1 Mpc in perpendicular offset. 

\subsection{Numerical Accuracy}\label{sec:tests}

The numerical action solution in this model has positions uniformly spaced in the expansion parameter $a(t)$, with $n_x=500$ steps between $a_1=0.1$ (meaning the initial redshift is  $z_i=9$) and the present at $a_{n_x+1}=1$. In the units of the computation (1~Mpc, 100~km~s$^{-1}$, and $10^{11}m_\odot$) relaxation terminates when the sum of the squares of the action gradients (eq.~[\ref{eq:eofm}]) of all particles at all time steps is SOS~$<10^{-18}$. The last iteration usually leaves the sum at SOS~$\la 10^{-20}$. This ensures that the effect on $\chi^2$ of a small change of a parameter signifies a change in the fit to the constraints rather than a change in the numerical error in the solution.

\begin{figure}[htpb]
\begin{center}
\includegraphics[angle=0,width=3.5in]{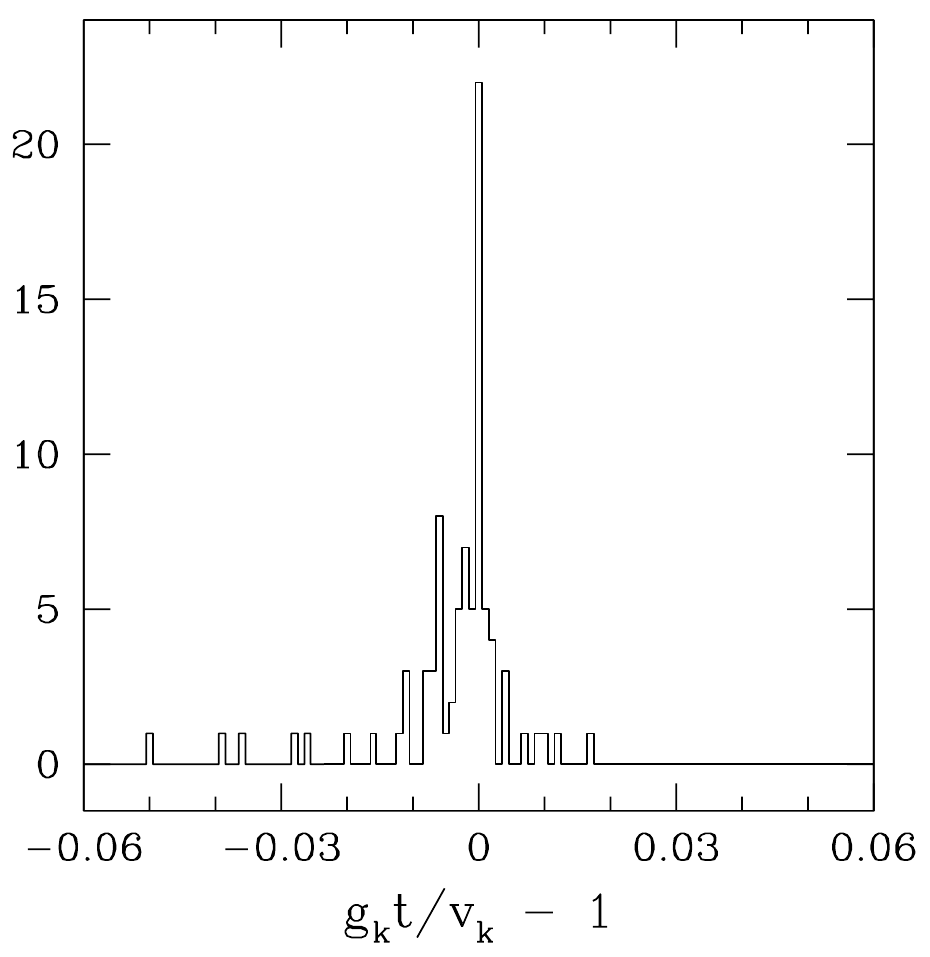} 
\caption{\small Distribution of fractional differences between the components $k=1,2,3$ of the initial LG galaxy peculiar velocities in linear perturbation theory and in the NAM solution.\label{Fig:gt}}
\end{center}
\end{figure}

The test of initial conditions in Figure~\ref{Fig:gt} compares the Cartesian components of the velocities at $1+z_i=10$ of the 28 LG galaxies in the NAM solution to the linear perturbation theory relation $\vec v=\vec gt$ in equation~(\ref{eq:gt}), where the peculiar acceleration $\vec g$ is computed from the positions at $z_i$ in the NAM solution. Most components agree to better than  than 0.5 percent, but there is a tail extending to $-5$\%\ for one component of the initial velocity of the Cassiopeia dwarf spheroidal galaxy. It will be recalled that this relation assumes the particle orbits satisfy the equation of motion back to $a(0)=0$,  with  coordinate displacements scaling as $a(t)$. NAM produces a solution in leapfrog approximation, and the first step, from $a=0$ to $a_1=0.1$, is a long one. The small scatter in Figure~\ref{Fig:gt} indicates that this first step has produced reasonably small --- though not entirely negligible --- errors in the initial condition in equation~(\ref{eq:gt}). That is, the approximation to the linear theory of growth of departures from homogeneity at large redshift is reasonably close, which ensures the wanted condition that peculiar velocities at  $1+z_i=10$ are growing in response to the peculiar gravitational acceleration that is causing the growth of structure. 

Numerical accuracy of the NAM solution from  $1+z_i=10$ to the present is checked by using initial positions and velocities derived from the solution for a numerical integration forward in time with 5000 steps uniformly spaced in $a(t)$, ten times the number in the NAM solution. Figure~\ref{Fig:gt} shows that the particle coordinate positions at $1+z_i\sim 10$ are changing in proportion to $a(t)$ to a good approximation, so initial conditions for the forward numerical integration are well approximated as
\beq
\vec x_{3/2}={\vec x_1+\vec x_2\over 2}, \quad  
\vec v_{3/2}= a_{3/2}\dot a_{3/2}{\vec x_2-\vec x_1\over a_2-a_1}.
\eeq
The positions $\vec x_1$ and $\vec x_2$ are from the first two steps in the NAM solution, the initial value of the expansion parameter is $a_{3/2}=(a_1+a_2)/2$, and $\dot a_{3/2}$ at $a_{3/2}$ is computed from the Friedmann equation. The largest differences between present positions and velocities in the forward integration and the NAM solution are 1.5~kpc, for NGC~6822, and 1.0~km~s$^{-1}$, for LMC. These are the orbits with the largest curvatures, at close passages of MW. The other differences of distances and redshifts generally are less than 10~pc and 0.01~km~s$^{-1}$. That is, we expect no problems with numerical errors.

\begin{figure}[htpb]
\begin{center}
\includegraphics[angle=0,width=5.5in]{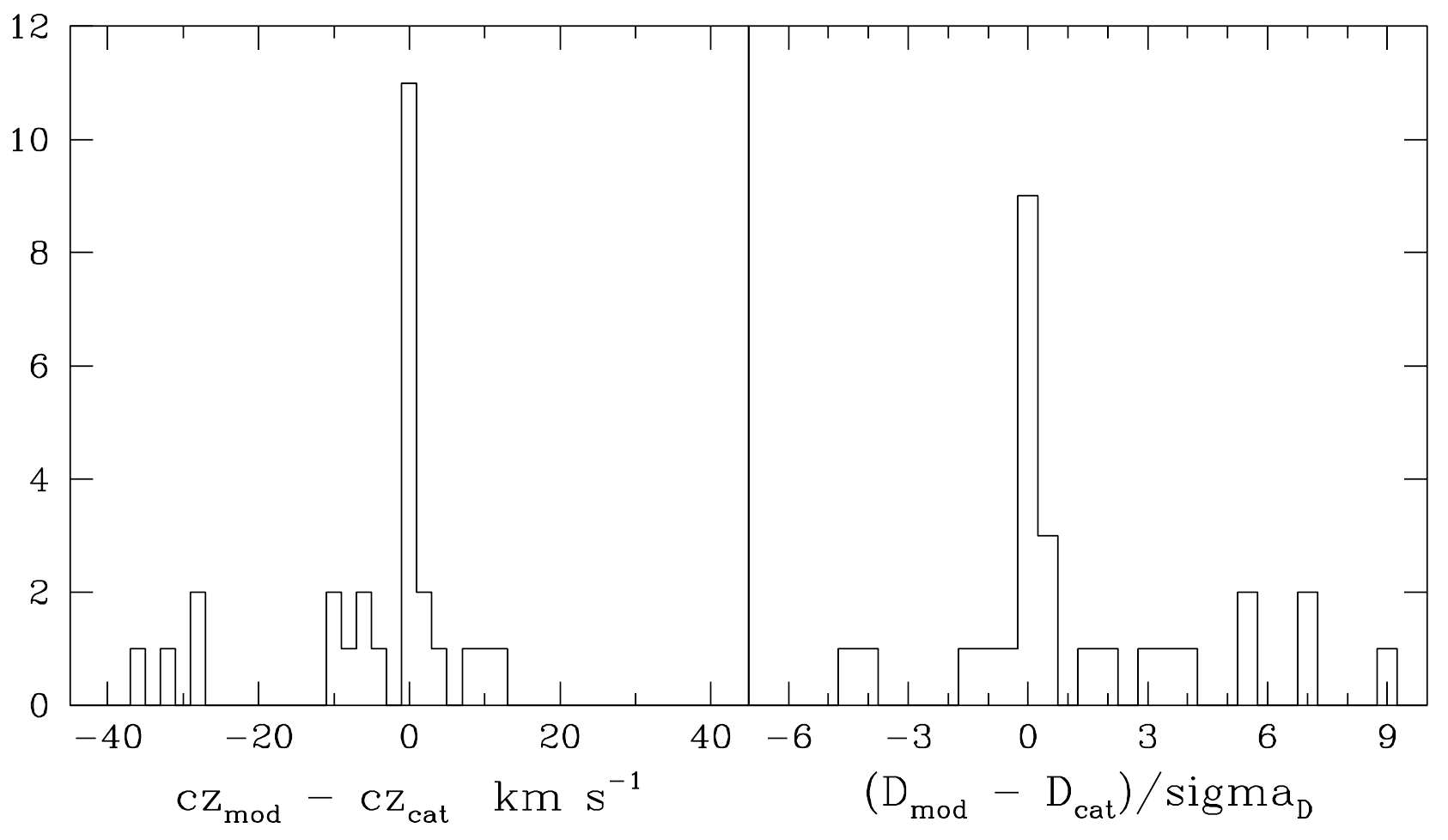} 
\caption{\small Differences between model and catalog LG galaxy redshifts and distances. The  differences of distances in the right-hand panel are divided by the catalog measurement uncertainties.\label{Fig:2}}
\end{center}
\end{figure}

\section{Numerical Results}\label{sec:results}

Table~1 lists the differences between LG galaxy model and measured redshifts, $\Delta cz = cz_{\rm mod}-cz_{\rm cat}$, and distances, $\Delta D = D_{\rm mod}-D_{\rm cat}$.  Figure~\ref{Fig:2} shows the distributions of these differences. The model redshifts of 23 of the 27 galaxies are within the goal of $10$~km~s$^{-1}$ differences from the catalog redshifts. The four exceptions ---  Cetus, Tucana, DDO~210, and the Sagittarius dwarf irregular --- have model redshifts that are too small by $\sim 30$~km~s$^{-1}$; they make up the left-hand tail of the redshift error distribution in Figure~\ref{Fig:2}. The model distances of these four misfit galaxies all are too large, by $(D_{\rm mod}-D_{\rm cat})/\sigma_D\sim 6$ to~9 (where $\sigma_D$ is the catalog distance uncertainty, not the larger allowance in our $\chi^2$ measure of fit). These are the four largest distance discrepancies; they dominate the right-hand tail in the distance error distribution in Figure~\ref{Fig:2}. The  next two greatest discrepancies relative to the catalog errors are $|D_{\rm mod}-D_{\rm cat}|/\sigma_D\simeq 5$ for IC10 and M33. The next largest are $-4.5$ for NGC 185, 3.8 for NGC 3109, and 3.6 for Sextans B. The rest are within three times the catalog uncertainty. 

The offsets $D_\perp$ of the particles from the catalog positions transverse to the line of sight are listed in Table~1. The offset of LMC is  $D_\perp=3$~kpc, which seems acceptable, and the rest are well within what seemed to be optimistic goals when the computation was planned. The last column in Table~1 is the velocities of the LG galaxies at $1+z_i=10$. They are comparable to reasonable-looking  escape velocities from small galaxies, and arguably in line with the near close packing of the galaxies at $z_i$. 

The conversion from model galactocentric to heliocentric velocities depends on the circular velocity of the local standard of rest. The $\chi^2$ minimization brings the circular velocity to 
\beq
v_c = 255.7\hbox{ km s}^{-1}. 
\label{eq:vc}
\eeq

\begin{table}[t]
\centering
\begin{tabular}{llrrrrrr}
\multicolumn{8}{c}{Table 4: Masses and Proper Motions}\\
\noalign{\medskip}
\tableline\tableline\noalign{\smallskip}
 & Name  & $\log(m)^{\rm a}$ & $m/L_K$ & ${v_\alpha}^{\rm b}$ & ${v_\delta}^{\rm b}$ &
 ${\mu_\alpha}^{\rm c}$ & ${\mu_\delta}^{\rm c}$ \\
   \noalign{\smallskip}
\tableline
\noalign{\smallskip}
   1&    MW&  12.19&   34& \nodata &\nodata&\nodata&\nodata \\
  2&   LMC&  10.68&   19&$  556.6$&$  118.4$&$  2.104$&$  0.448$\\
  3& Leo I&   7.30&   43&$   69.5$&$ -369.3$&$  0.061$&$ -0.323$\\
  4& Leo T&   7.45&   49&$  -66.8$&$ -255.0$&$ -0.034$&$ -0.130$\\
  5&   Phx&   6.90&   48&$  193.1$&$ -204.7$&$  0.088$&$ -0.093$\\
  6& N6822&   9.85&   59&$  -87.1$&$ -208.8$&$ -0.037$&$ -0.088$\\
  7&  N185&  10.58&  179&$  205.4$&$ -227.8$&$  0.086$&$ -0.095$\\
  8&  LGS3&   5.60&   52&$  141.8$&$   29.6$&$  0.044$&$  0.009$\\
  9&Cet dS&   6.90&   50&$  206.9$&$ -183.4$&$  0.046$&$ -0.041$\\
 10&  N147&   9.40&   16&$  194.1$&$ -374.9$&$  0.054$&$ -0.105$\\
 11& Leo A&   7.60&   50&$  -60.6$&$ -344.9$&$ -0.017$&$ -0.097$\\
 12& I1613&   9.66&   80&$  165.5$&$ -141.0$&$  0.046$&$ -0.039$\\
 13&  AXIV&   7.00&   49&$  338.3$&$ -301.0$&$  0.096$&$ -0.086$\\
 14&Cas dS&   7.78&   50&$  320.9$&$ -178.2$&$  0.085$&$ -0.047$\\
 15&  IC10&  10.24&   20&$   31.9$&$  -16.5$&$  0.007$&$ -0.003$\\
 16&   M31&  12.39&   45&$  240.5$&$  -63.1$&$  0.061$&$ -0.016$\\
 17&  AXII&   6.60&   50&$  198.9$&$ -177.5$&$  0.043$&$ -0.039$\\
 18&   M33&  10.88&   18&$   52.0$&$   20.8$&$  0.016$&$  0.006$\\
 19&Tucana&   6.30&   51&$  101.0$&$ -306.3$&$  0.016$&$ -0.047$\\
 20&PegDIG&   7.72&   56&$  137.4$&$  -80.2$&$  0.033$&$ -0.019$\\
 21&DDO210&   7.70&   54&$  -21.5$&$ -267.0$&$ -0.003$&$ -0.040$\\
 22&   WLM&   8.98&   35&$  204.0$&$ -210.5$&$  0.044$&$ -0.046$\\
 23&SagDIG&   7.68&   52&$ -114.6$&$ -285.5$&$ -0.017$&$ -0.043$\\
 24& N3109&   9.51&   90&$   13.3$&$ -117.4$&$  0.002$&$ -0.015$\\
 25&Antila&   6.64&   50&$   20.9$&$  -97.0$&$  0.003$&$ -0.015$\\
 26& U4879&   8.24&   48&$  -20.8$&$ -333.7$&$ -0.003$&$ -0.053$\\
 27& Sex B&   9.14&   51&$  -29.6$&$ -290.1$&$ -0.004$&$ -0.036$\\
 28& Sex A&   9.11&   49&$  -35.4$&$ -251.8$&$ -0.005$&$ -0.034$\\
 \noalign{\smallskip}
\tableline
\noalign{\smallskip}
\multicolumn{8}{l}{$^{\rm a}$Solar masses\ \  $^{\rm b}$km s$^{-1}$\ \  $^{\rm c}$mas yr$^{-1}$ } \\
\end{tabular}
\end{table}

Table~2 compares measured and model proper motions of LMC, M33 and IC10. The largest discrepancy is the motion of IC10 in declination, at three times the measurement uncertainty. This is arguably acceptable within the approximations of the model, and considering the difficulty of the measurement. Table~4 lists heliocentric velocities normal to the line of sight in the directions of increasing right ascension and declination, and the corresponding angular velocities, for all the LG galaxies.

The mass-to-light ratios $m/L_K$ in Table~4 are based on the 2MASS K-band absolute magnitudes $L_{\rm K}$  in the Local Universe Catalog. These magnitudes are reliable for the larger galaxies, quite uncertain for the least luminous ones. The smaller galaxies were arbitrarily assigned nominal masses for the computation of $\chi^2$ derived from $m/L_{\rm K}=50$. Where that mass is too small to affect the solution the final mass nevertheless is changed by the insistence of the computation on small but nonzero trial mass shifts in the approach to a minimum of $\chi^2$, but the change is small. The $\sim 19$ final values $m/L_{\rm K}\sim 50$  signify masses that seem to be too small to matter. The possible significance of the curious values of some of the other masses is considered in Section~\ref{sec:discussion}. 

\begin{figure}[t]
\begin{center}
\includegraphics[angle=0,width=4.5in]{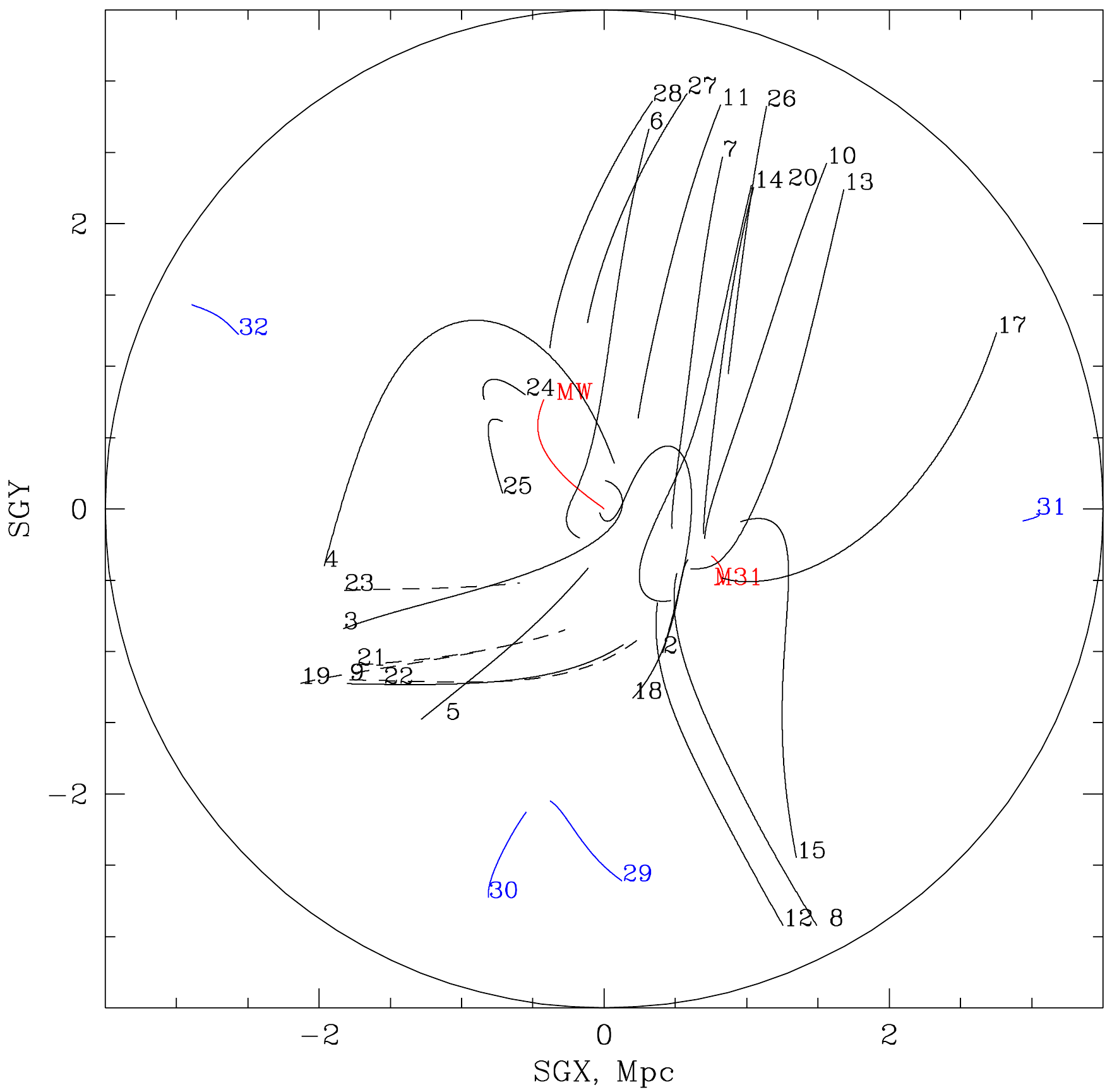} 
\caption{\small Model orbits in comoving supergalactic coordinates. The numbers are keyed to the names in Tables~1, 3 and~4. The blue curves show the external actors. The dashed curves for the four misfit galaxies illustrate their apparently common origin.\label{Fig:3}}
\end{center}
\end{figure}

\begin{figure}[htpb]
\begin{center}
\includegraphics[angle=0,width=4.5in]{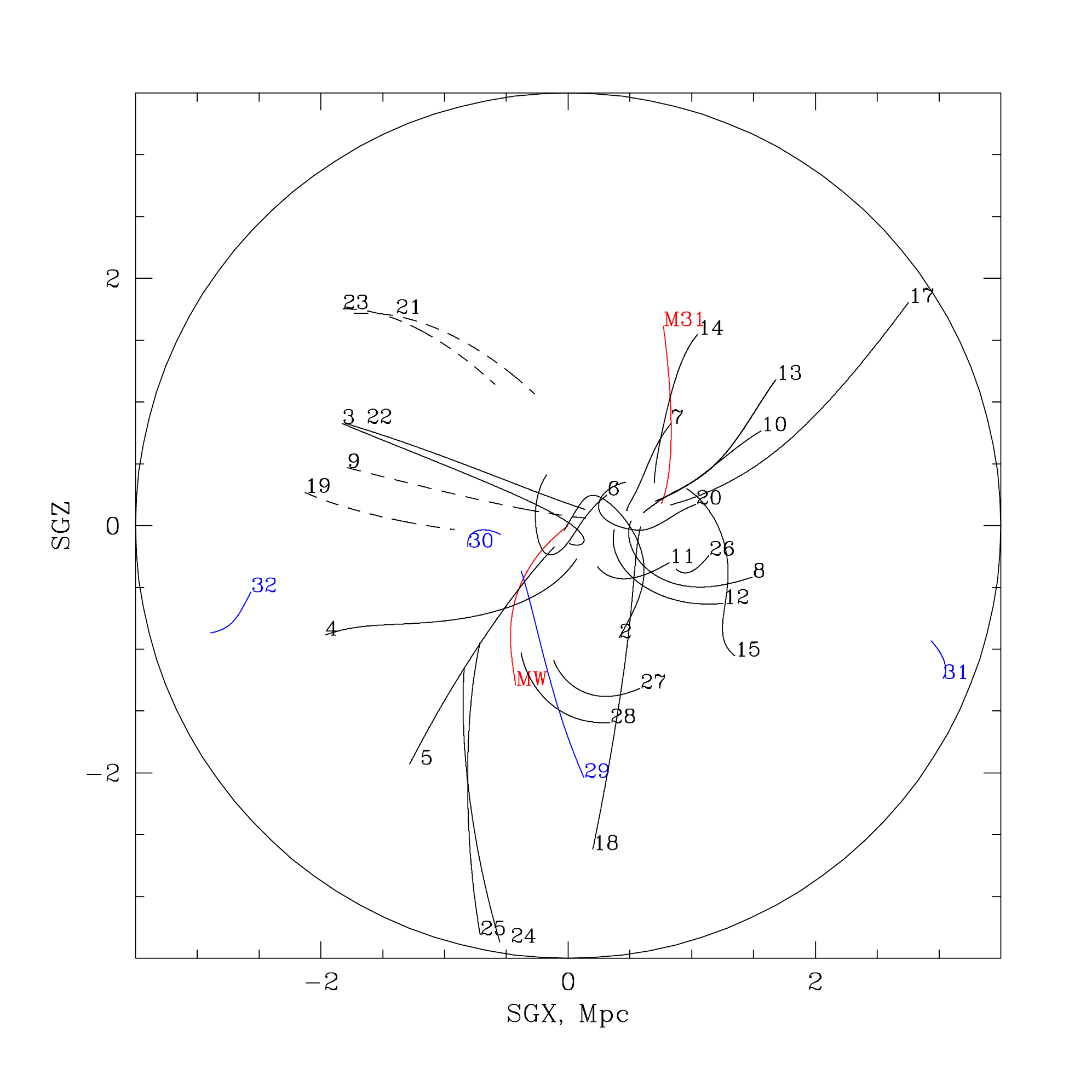} 
\caption{\small The same as Fig.~\ref{Fig:3} for the orthogonal view toward $SGL = 90^\circ$, $SGB = 0$.\label{Fig:4}}
\end{center}
\end{figure}

\begin{figure}[htpb]
\begin{center}
\includegraphics[angle=0,width=4.5in]{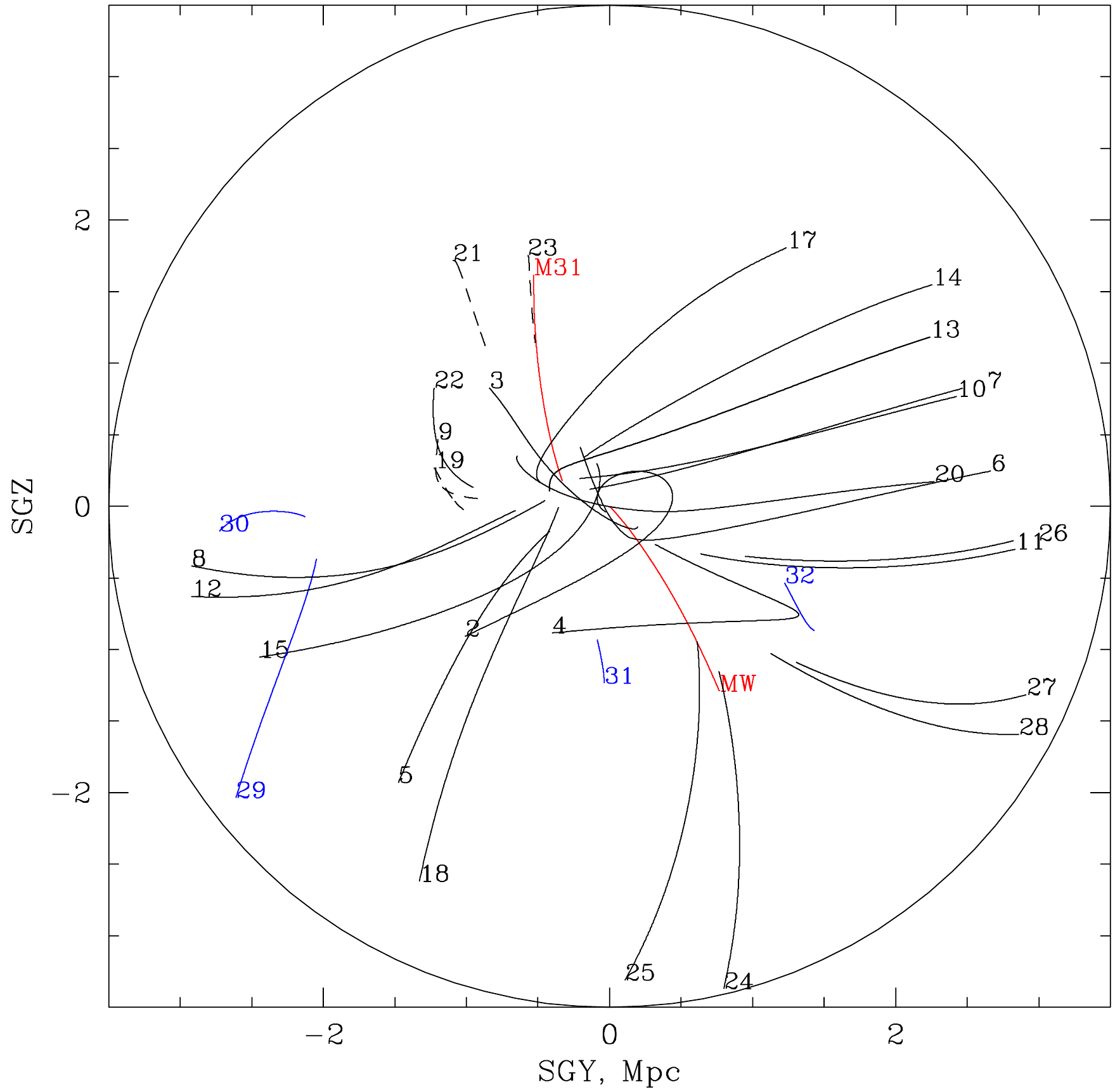} 
\caption{\small The same as Fig.~\ref{Fig:3} for the orthogonal view from $SGL = 0$, $SGB = 0$.\label{Fig:5}}
\end{center}
\end{figure}

Figures~\ref{Fig:3} to~\ref{Fig:5} show orthogonal views of the model orbits plotted in comoving supergalactic coordinates. The center of mass of the LG galaxies plus the external actors is at rest and the origin is at the present position of MW. The labels near  initial positions at expansion factor $1+z_i=10$ are keyed to names in the first two columns in Tables~1, 3, and~4. The orbits of the two dominant Local Group galaxies, MW and M31, are plotted in red, and the four external actors that are supposed to give a phenomenological description of the effect of mass outside $1.5$~kpc are shown in blue. The orbits of the four misfit galaxies with the poorest fits to catalog redshifts and distances are plotted as the dashed black curves, and the orbits of the other LG galaxies are shown as the solid black curves.

\section{Discussion}\label{sec:discussion} 

The focus of this dynamical study has been on fitting the model to measured positions and velocities of the 27 galaxies at distances less than 1.5~Mpc.  We emphasize again that the measures of the external actors in Table~3  are chosen for a phenomenological description of the effect of external mass on the motions of the LG galaxies implied by the pattern of LG velocities and positions; the parameters of the external actors should not be taken as meaningful estimates of the properties of the named systems. We do take it to be an  interesting project to attempt to interpret the large distance offsets normal to the line of sight for the actors named Maffei and Centaurus, and their curiously small masses, as an indication of the nature of the external mass distribution, but that is not attempted here. 

Since the LG galaxy masses are left nearly free to aid the fit we count as free parameters the ten LG galaxy masses that seem to be large enough to matter. The redshifts and distances of the external actors are close to their assigned values, meaning they do not seem to matter, but we count as significant parameters the four perpendicular offsets and four masses of the actors.  
This adds up to 18 free parameters that have yielded tolerably good fits  to six components of proper motion, 23 redshifts, and 18 distances. Whether the four redshifts off by $-30$~km~s$^{-1}$ and nine distances off by about $400$~kpc are to be counted as modest succeses or serious failures is a matter of judgement.  To the count of successful fits to constraints should be added the $28\times 3$ conditions on the components of the initial peculiar velocities in equation~(\ref{eq:gt}), as illustrated in Figure~\ref{Fig:gt}, and the 28 initial velocities in the eighth column in Table~1,  which are judged to be reasonably consistent with the close packing of protogalaxies at $1+z_i=10$. The count of constraints on initial conditions is not well defined --- we can only offer estimates of reasonable initial velocities, and equation~(\ref{eq:gt}) is an approximation --- but it adds to the case that the model is a useful approximation, albeit one that needs improvement. 

Of the 13 departures from tolerable fits to redshifts and distances, eight belong to four misfit galaxies, Cetus, Tucana, DDO~210, and the Sagittarius dwarf irregular. They have similar discrepancies in redshifts and distances, and they have similar orbits (plotted as the dashed lines in Figs.~\ref{Fig:3} to~\ref{Fig:5}) that  emanate from $SGL\sim 200^\circ$, $SGB\sim 30^\circ$. It may be significant that this is in the direction of the Local Void (Tully et al. 2008). The common features --- low redshifts, large distances, and similar orbits --- argue against the idea that measurement errors are to blame, and invites the speculation that they are manifestations of an inadequate external mass model that would have to be particularly serious near these four galaxies. If an adjustment of our phenomenological representation of the external mass could reduce the peculiar gravitational acceleration toward MW near the four misfits it would allow larger redshifts at lower present distances, in the direction wanted to improve the fit. A search for a fifth external mass capable of producing this effect has not yielded anything promising, however. An explanation of the enigmatic properties of these four misfit galaxies  remains an interesting open issue. 

After the four misfits the next two largest numerical anomalies are the differences between model and catalog distances relative to catalog uncertainties for M33 and IC10. It is not unreasonable to expect that some distance measurements have errors well outside the catalog uncertainties, and the low galactic latitude of IC10 may make this distance measurement particularly problematic, though both Sanna et al (2008) and Kim et al. (2009) find distances to IC10 that are short of the model value in Table~1. The orbits of M33 and IC10 are strongly constrained by  measurements of their proper motions. These proper motions require corrections for the motions of the masers relative to the host galaxies. We can suggest no problem with this carefully done correction, but additional considerations of these valuable constraints on the orbits of M33 and IC10 and on the dynamics of the Local Group certainly will be welcome.

Although the orbit of M33 fits demanding constraints from its redshift and proper motion as well as the initial conditions, the result is  problematic. The weight of the evidence informing the Local Universe Catalog is that M33 is more distant than M31, but the model puts M33 closer. Also, the model has M33 approaching M31 for the first time since high redshift, while the evidence is that these two galaxies passed close enough to have disturbed the H{\small I} disk of M33 and drawn out stellar and H{\small I} streams possibly connecting M33 to M31  (Rogstad et al. 1976; McConnachie et al. 2009; Putman et al. 2009;  Richardson et al. 2011).  To be considered is that some galaxies that are quite isolated from visible companions nevertheless have apparently disturbed H{\small I} disks (Kreckel et al. 2011), and that models for stellar halos can produce stellar streams that originated at high redshift (Cooper et al. 2010; Wang, Peebles, and Nusser 2011), when the galaxies were closer together. Separating these phenomena from the possible effect of a more recent interaction of M31 and M33 is an interesting challenge for future work. 

The model masses have problematic features, including the exceptionally large mass-to-light ratio of NGC~185, and, since the model would have similar circular velocities in MW and M31, the dissimilar model masses. Assessment of issues with masses must take account of three points. First, we have useful statistical measures of the relation between total galaxy masses and the stellar masses and velocities, but the scatter in the relation for individual galaxies is not at all well measured: it is to  be explored by dynamical analyses such as this. Second, our measure of the fit of the model to the constraints uses nominal values of the galaxy masses that were thought to be particularly important for the dynamics. The relaxation to a minimum of the $\chi^2$ measure of fit can end up at a local minimum closer to the nominal mass than another local minimum at a more accurate mass. The effect is real even though the penalty of a mass adjustment is low.  An example is the small model mass of LGS~3, only 10\% of the  stellar mass found by Hidalgo et al. (2011). This may be a result of the nominal choice of a mass small enough not to matter. Another example is the choice of nominal masses of  MW and M31, $1.5\times 10^{12}m_\odot$ and $3\times 10^{12}m_\odot$. This approximates conventional wisdom, but the choice may have led to a local minimum of $\chi^2$ with similar model masses, $1.6\times 10^{12}m_\odot$ for MW and  $2.4\times 10^{12}m_\odot$ for M31. An attempt to adjust the solution to bring these two masses closer by adding to $\chi^2$ a penalty for a significant mass difference had little effect. This might be expected because changing the masses of these two galaxies requires consistent adjustments of the redshifts and distances of many LG galaxies, a slow operation by the present numerical method. The third point to consider is that our analysis has allowed masses considerable freedom to float to aid the fit to distances and velocities. This means that some erroneous choices of orbits may have been be made to fit the measured redshifts and distances by the choice of erroneous masses. Further investigation of the last two issues will require reconstruction of the model, which we may hope will be aided by future still tighter constraints on distances and proper motions that reduce the chance of including erroneous orbits. 
 
The fit to constraints on LG galaxy positions and velocities relaxes the MW circular velocity to  $v_c = 256$~km~s$^{-1}$. This is 36 km s$^{-1}$ larger than a conventional value and 26 km s$^{-1}$ larger the nominal central value used in the computation of $\chi^2$,  but it is consistent with the Reid et al. (2009a) measurement, $254 \pm 16$~km~s$^{-1}$. We cannot assign a meaningful uncertainty to this or the other numerical results from the dynamical model solution because our $\chi^2$ measure of fit is dominated by a few large discrepancies, and some (such as the four misfit galaxies in Figs.~\ref{Fig:3} to~\ref{Fig:5}) certainly represent systematic errors in the model. That is, our $\chi^2$ measure has no formal significance. We can point out that the model value was $v_c=241$~km~s$^{-1}$ in the solution with 10 LG galaxies (plus the 4 external actors) and $v_c$ increased as more distant galaxies were added to the solution. That is, the larger value of $v_c$ in the final model changed the relation between heliocentric and galactocentric redshifts in the direction wanted to aid the fit to the redshifts of more distant LG galaxies. This means that if $v_c$ proved to be close to the standard value, $220$~km~s$^{-1}$, the challenge to the model likely would not be related to the problematic orbits of the nearer galaxies M33 and IC10.

The computation allows the present positions of the particles to be offset from the observed positions of the galaxies on two considerations,  that the approximate model should be allowed some  error in this direction and that this affords an indication of whether the dark matter around a galaxy may be significantly displaced from the stars. The generally small values of the offsets $D_\perp$ in Table~1 mean we have no evidence that the stars are not well centered on the dark matter, but tighter constraints on the model errors are needed for a significant exploration  of this important issue.
 
\begin{figure}[t]
\begin{center}
\includegraphics[angle=0,width=6.0in]{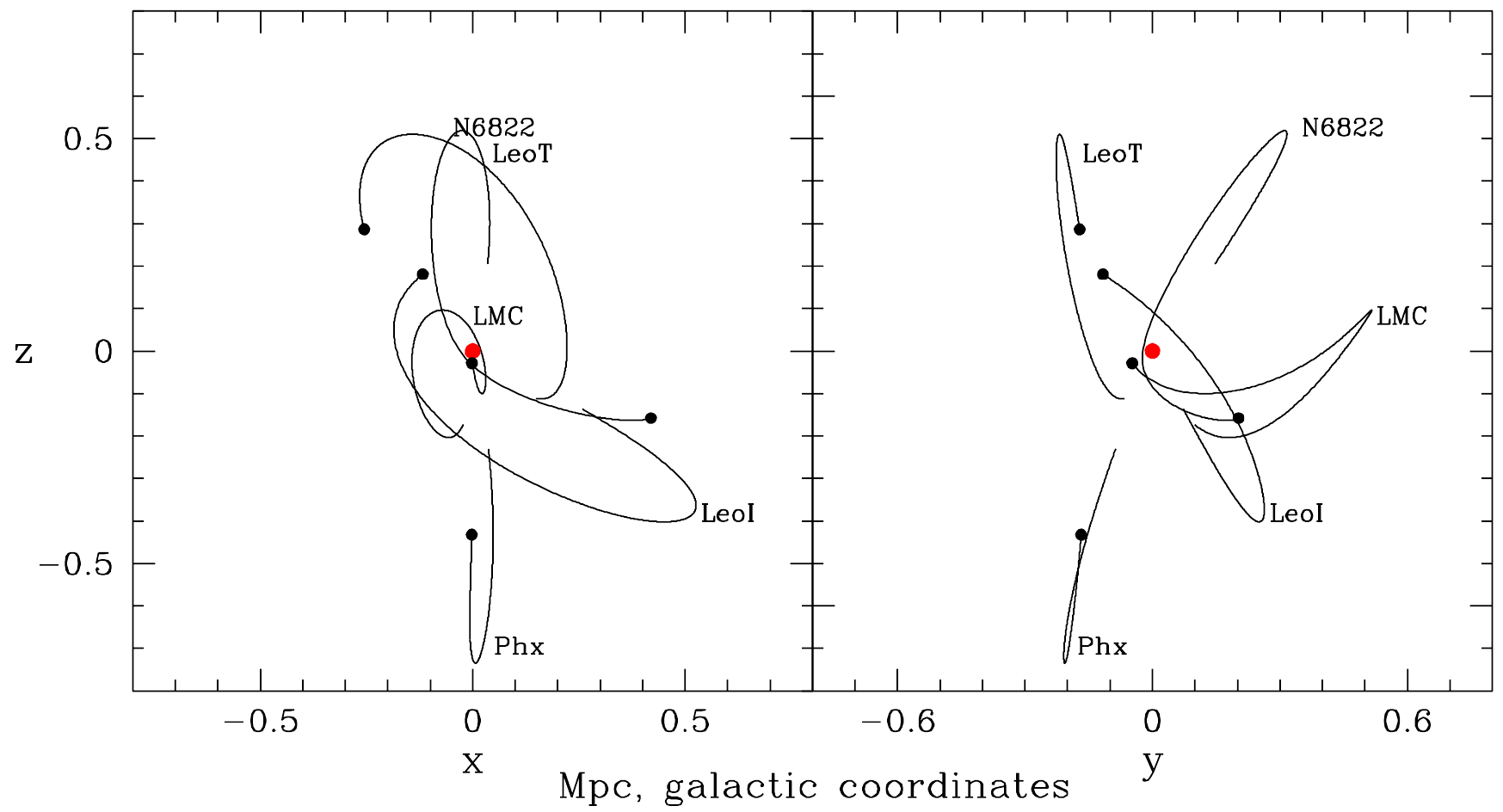} 
\caption{\small Orbits relative to MW, at the red circle, of the five closest neighbors.\label{Fig:orbits}}
\end{center}
\end{figure}

\begin{table}[htpb]
\centering
\begin{tabular}{lrrrr}
\multicolumn{5}{c}{Table 5: Nearer MW Neighbors}\\
\noalign{\medskip}
\tableline\tableline\noalign{\smallskip}
 Name & ${d_{\rm max}}^{\rm a}$ & $a_{\rm max}$  & ${d_{\rm min}}^{\rm a}$ & $a_{\rm min}$ \\
   \noalign{\smallskip}
\tableline
\noalign{\smallskip}
LMC  & 530 & 0.59  & 56    & 1.00 \\
 Leo I  & 680 & 0.43 &   200 & 0.93  \\
Leo T  & 580 & 0.71 &  420 & 1.00  \\
Phoenix & 760 & 0.60 & 460& 1.00  \\
N~6822  & 610 & 0.43 &  34 & 0.79 \\
 \noalign{\smallskip}
\tableline
\noalign{\smallskip}
\multicolumn{5}{l}{$^{\rm a}$kpc\qquad } \\
\end{tabular}
\end{table}

Figure~\ref{Fig:orbits} shows the evolution of positions relative to MW of the five galaxies now closest to MW. The coordinates are galactic and the lengths are physical. Positions at redshift $z=1$ are near the galaxy labels and  present positions are at the filled black circles. None of the five has completed very much more than one orbit. The second and third columns in Table~5 list the maximum distances $d_{\rm max}$ from MW and the expansion parameters $a_{\rm max}$ (referred to $a=1$ at $z = 0$) at $d_{\rm max}$ for the five LG galaxies in Figure~\ref{Fig:orbits}. The next two columns list the minimum distances of these galaxies at $a>a_{\rm max}$ and the expansion parameter at this minimum distance. The model orbits of the nearest satellites of M31 relative to its position have similar appearance to Figure~\ref{Fig:orbits}, though none pass as close to M31 as do LMC and NGC~6822 to MW. This should be taken with caution, however, because the  present distances from MW to its nearest neighbors are reasonably well known, while modest relative distance errors could introduce considerable errors in the present distances from M31 of its nearer neighbors.

In the model, the galaxies Leo T and Phoenix are approaching MW for the first time and  LMC is  now passing MW. The galaxy Leo~I is moving away after closest approach of 200~kpc at redshift $z=0.08$. Leo~I has been compared to closer dwarf spheroidal satellites of MW that have completed many orbits (and for this reason are not in the compuation) and in the process may have been expected to have lost much of their gas. Leo I does not have detected H{\rm I} (Young 2000), despite the single relatively distant passage at 200~kpc distance. If the model orbit is right it means either that the ram pressure of plasma clouds at this distance was capable of removing the H{\small I}, or that Leo~I was capable of driving away the remnant H{\small I} by itself. The predicted much closer passage of NGC\,6822,  34 kpc at redshift $z=0.27$, left H{\small I} around this galaxy, and might be expected to have left a tidal tail similar to the Magellanic Stream, though perhaps significantly dissipated by the passage of time. It may be significant that the redshift in the H{\small I} around NGC~6822 varies almost linearly with position along the long axis of the H{\small I} distribution, not the  expected behavior of a rotationally-supported disk in a standard dark matter halo (Weldrake, de Blok,  \& Walter  2003). To be investigated is whether the observed distribution and motion of the H{\small I} and stars in this galaxy might be consistent with remnant effects of a close MW passage. 

Table~4 lists   proper motions of the LG galaxies. Van der Marel \& Guhathakurta (2008) deduce from the motions of its satellites that the transverse components of the heliocentric velocity of M31 are $v_\alpha=78\pm 41$ km~s$^{-1}$ and $v_\delta=-38\pm 34$ km~s$^{-1}$. The former differs from the dynamical model by four times the stated uncertainty and the latter agrees with the model within the uncertainty. Since the model has its own uncertainty, though we have no meaningful estimate of it, the two measures of the heliocentric velocity of M31 seem to be tolerably consistent. 

Measurements of proper motions of the nearer galaxies in the Local Universe Catalog would be particularly important additions to the constraints on local dynamics because their distances from MW are fairly well  known, and their model orbits reach far enough from MW to be to be sensitive to the evolution of the local mass distribution on scales of a few megaparsecs. The dwarf elliptical Leo I may be an interesting candidate for an optical measurement of the mean angular velocity of the stars. In the model it has the second largest heliocentric angular velocity, 0.33~mas~yr$^{-1}$, about  three times the uncertainty in the measurement of the angular velocity of LMC. Half of the  angular velocities of LG galaxies in Table~4 are greater than $0.05$~mas~yr$^{-1}$, well above the detected  motions of masers in IC10 and M33 (Brunthaler et al. 2005, 2007). Depending on what searches for masers in LG galaxies reveal, measurements of angular motions by present VLBI technology has the potential to open a new frontier in constraints on local dynamics. We may hope this frontier will be opened far more widely by new advances in the optical (the  GAIA Science Mission) and VLBI (Reid et al. 2009b).

\begin{figure}[htpb]
\begin{center}
\includegraphics[angle=0,width=6.0in]{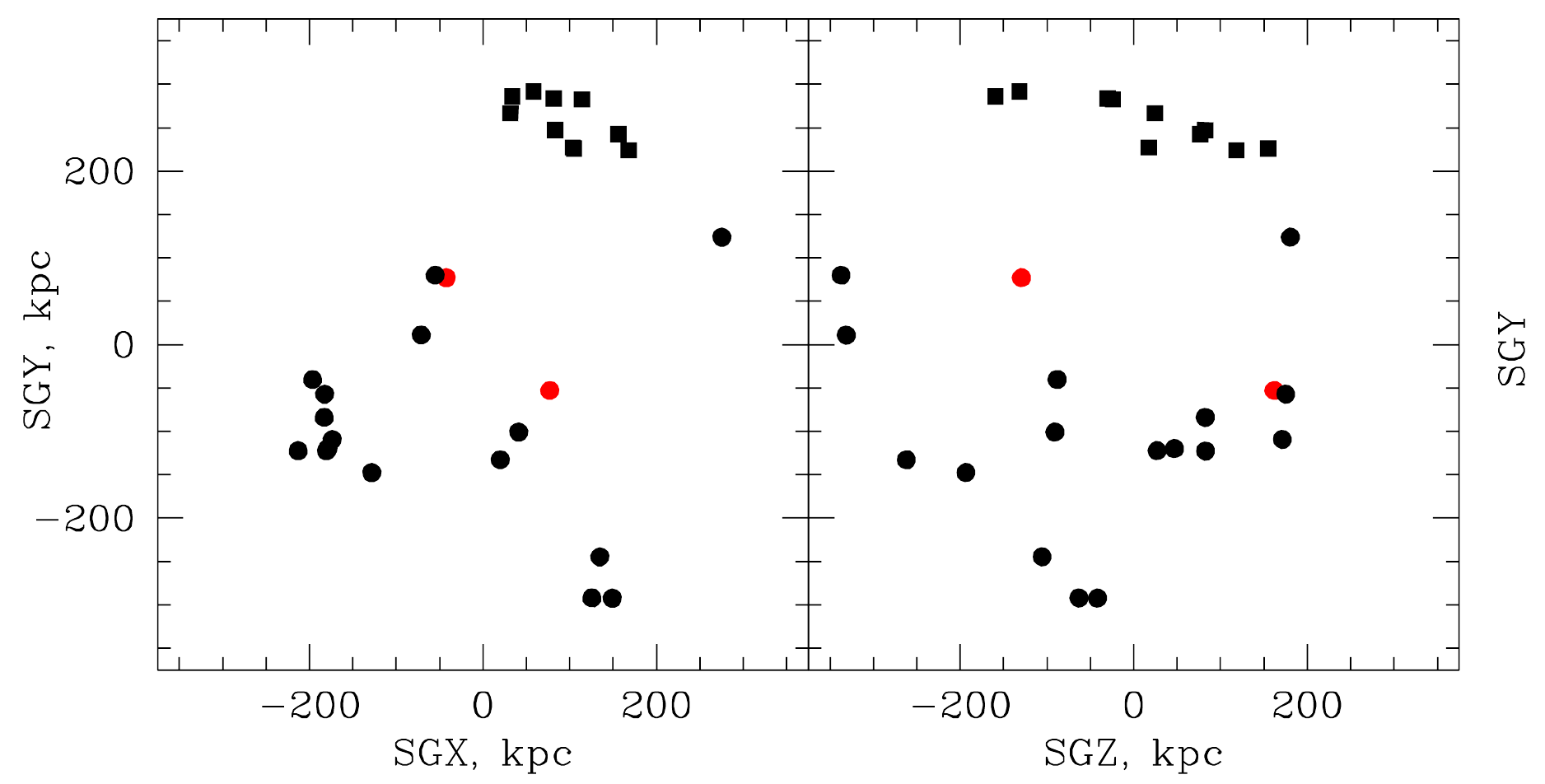} 
\caption{\small Positions of the LG galaxies at $1+z_i=10$ in physical supergalactic coordinates. The ten galaxies in the concentration at SGL$\sim70^\circ$, SGB$\sim0$ are plotted as filled squares.\label{Fig:7}}
\end{center}
\end{figure}

A group of ten LG galaxies emanate from high redshift along similar paths. This is seen  in Figures~\ref{Fig:3} to~\ref{Fig:5}, and it is illustrated in another way in Figure~\ref{Fig:7}, which shows physical positions at the starting time of the computation, at expansion factor $1+z_i=10$. The external actors are not shown in this figure, the initial positions of MW and M31 are plotted in red, and the LG galaxies not in the group of ten are shown as filled circles. The ten with similar initial positions, shown as filled squares, are in order of increasing present distance from MW the galaxies NGC~6822, NGC~185, NGC~147, Leo~A, Andromeda~XIV, Cassiopeia dwarf spheroidal, Pegasis dwarf, UGC~4879, and Sextans~A and~B. One might include a more distant eleventh member of the group, Andromeda~XII. It is plotted as the filled circle at largest SGX and SGY in Figure~\ref{Fig:7}, and has the label 17 in Figs.~\ref{Fig:3} to~\ref{Fig:5}. The initial positions of the inner ten are in a band of physical length $\sim 400$~kpc that is close to perpendicular to the supergalactic plane, and centered at SGL~$\sim 70^\circ$, SGB$\sim0$. The narrow dimension is less than $75$~kpc. If this distinct early concentration had been a little tighter the group could have merged into a  protogalaxy. It might prove interesting to check whether these ten parts of a possible failed protogalaxy have more in common among themselves than among the other low mass LG members, perhaps in the H{\small I} mass fraction or the distributions of metallicity or stellar evolution ages, reflecting a near common origin. 

\section{Concluding Remarks}

This model for the Local Group certainly has problems. Perhaps most serious is the orbit of M33, which fits the measured distance, redshift and proper motion, within tolerable errors, along with the initial condition, but not the indications that M33 passed close to M31 at modest redshift. This situation certainly requires further examination. The common anomalies in the redshifts and distances of the four misfits whose near common orbits are plotted as dashed lines in Figures~\ref{Fig:3} to~\ref{Fig:5} seem to be pointing to an error in the simplified phenomenological description of the effect of the mass distribution outside the Local Group. Investigating that will require a more detailed dynamical analysis that takes explicit account of where the larger external masses are observed to be now and where they are computed to have been. We hope to report on an exploration of this analysis in due course. The values of some of the masses derived from the dynamics are curious, and may be in part the result of compensation for systematic errors in the model. 

The very real problems should not obscure the evidence that this dynamical model has proved to be a useful approximation to reality, from the success in fitting the considerable number of observational constraints on positions and velocities in Tables~1 and~2, and perhaps also the Milky Way circular velocity $v_c$ in equation~(\ref{eq:vc}), with orbits that have the initial conditions required by a reasonable approximation to the cosmological growth of structure. It is notable also that the model has led to the considerable list of issues for further consideration presented in Section~\ref{sec:discussion}. 

In the past decade advances in detections of nearby galaxies, and in the measurements of galaxy distances and velocities, have greatly enriched the study of dynamics in the now densely sampled region within 1.5~Mpc distance. We have the prospect of another advance of the same order from work in progress, or that seems to be feasible, on measurements of galaxy distances and proper motions. That will present us with a really serious but fascinating challenge for the development of numerical methods of analysis of a tangled web of constraints on what happened in the local universe.

\acknowledgments

We are grateful to Mark Reid and Jeremy Darling for guidance to prospects for measurements of proper motions of Local Group galaxies.

\appendix
\section{Yet another Version of the Numerical Action Method}

This version produces a solution to the equation of motion in leapfrog approximation by the joint relaxation of all coordinatess toward a stationary point of the action in the direction indicated by the first and second derivatives of the action. This  joint relaxation  of all coordinates describing all orbits  considerably speeds the computation when applied to a modest number of particles. 

In the cosmologically flat universe of this analysis the expansion parameter satisfies 
\beq
{\dot a^2\over a^2} = {H_o^2\Omega\over a^3} + (1 - \Omega)H_o^2,\qquad
{\ddot a\over a} = -{H_o^2\Omega\over 2a^3} + (1 - \Omega)H_o^2,
\eeq
with present value $a_o=1$. 

Particle orbits are represented by comoving positions $x_{i,k,n}$ for the particle label $1\leq  i\leq n_p$, cartesian component $1\leq k\leq 3$, and time step $n\leq n_{x+1}$. The earliest computed positions are at $a_1=a_i$, the given present positions are at $a_{x+1}=1$, and the coordinates to be adjusted are at $1\leq n\leq n_x$. The leapfrog numerical integration commences at $a_{1/2}=0$ at the nominal singular start of expansion.

The equation of motion in leapfrog approximation is 
\beqa
0 = S_{i,k,n} &=& -{a_{n+1/2}^2\dot a_{n+1/2}\over a_{n+1} - a_n}(x_{i,k,n+1}-x_{i,k,n}) \nonumber\\
&+& {t_{n+1/2}-t_{n-1/2}\over a_n} \left[{\cal G}_{i,k,n} +
{1\over 2}\Omega H_o^2 x_{i,k,n}\right]\label{eq:eofm}\\
&+& {a_{n-1/2}^2\dot a_{n-1/2}\over a_{n} - a_{n-1}}(x_{i,k,n}-x_{i,k,n-1}). \nonumber
\eeqa
The expansion time $t_{n+1/2}$ is computed from the analytic expression for the time from $a_0=0$ to $a_{n+1/2}=(a_n+a_{n+1})/2$. The physical acceleration of particle $i$ produced  by the gravitational attraction of the other particles is ${\cal G}_{i,k,n}/a_n^2$.  The physical acceleration corresponding to the counter term in the square brackets may be written as
\beq
{\Omega H_o^2 x_{i,k,n}\over 2a_n^2}={4\pi\over 3}G\rho_n r_{i,k,n},
\eeq
where $r=ax$ is the physical coordinate and $\rho_n$ is the cosmic mean mass density at $a_n$. The counter term thus causes the peculiar acceleration to vanish when the physical acceleration matches that of a homogeneous universe. When the particles are represented as point masses,
\beq
{\cal G}_{i,k,n}= \sum_{j\not= i} Gm_j{(x_{j,k,n}-x_{i,k,n})\over |x_{i,n}-x_{j,n}|^3}. \label{eq:pointlike}
\eeq
In the limiting isothermal sphere model for the mass distribution in MW, $i=1$, when there is a nearby smaller galaxy, $j$, the terms for this pair in the sums in equation~(\ref{eq:pointlike}) are replaced with 
\beq
\delta {\cal G}_{j,k,n}= Gm_1{(x_{1,k,n}-x_{j,k,n})\over x_o|x_{1,n}-x_{j,n}|^2}, \qquad
\delta {\cal G}_{1,k,n}= Gm_j{(x_{j,k,n}-x_{1,k,n})\over x_o|x_{1,n}-x_{j,n}|^2},
\eeq
at separation 
\beq
|x_{1,n}-x_{j,n}| < x_o = G m_1/ (v_c^2a_n). 
\eeq
This means the physical acceleration of particle $j$ caused by MW is $v_c^2/r$, and the acceleration of MW caused by particle $j$ is $m_j/m_1$ times $v_c^2/r$, which conserves momentum.

Equation (\ref{eq:eofm}) is simplified by setting
\beq
F^+_n=  
{a_{n+1/2}^2\dot a_{n+1/2}\over a_{n+1} - a_n}, 
\quad
 F^-_n=  
{a_{n-1/2}^2\dot a_{n-1/2}\over a_{n} - a_{n-1}}
= F^+_{n-1}, \quad {dt_n\over a_n} = {t_{n+1/2}-t_{n-1/2}\over a_n},
\label{eq:F}
\eeq
where
\beq
F^-_1 = 0=F^+_0  \label{eq:bdyconds}
\eeq
follows from the conditions $a_{1/2}=0$ and $a\propto t^{2/3}$ at $a\rightarrow 0$.
This brings eq. (\ref{eq:eofm}) to
\beq
S_{i,k,n} = -F^+_n(x_{i,k,n+1}-x_{i,k,n}) +  F^-_n(x_{i,k,n}-x_{i,k,n-1})
+ {dt_n\over a_n}\left[{\cal G}_{i,k,n}  + 
{1\over 2}\Omega H_o^2 x_{i,k,n}\right].\label{eq:eofm1}
\eeq

The coordinates are relaxed toward a solution at $S_{i,k,n}=0$ by the coordinate shift $\delta x_{i,k,n}$ that satisfies
\beq
S_{i,k,n} + \sum_{j,k',n'}S_{i,k,n;j,k',n'}\delta x_{j,k',n'} =0. \label{eq:shifts}
\eeq
In this expression the nonzero derivatives of the $S_{i,k,n}$ with respect to the coordinates are
\beqa
S_{i,k,n;i,k,n+1} &=& - F^+_n, \qquad S_{i,k,n;i,k,n-1} = - F^-_n, \nonumber\\
 S_{i,k,n;j,k',n} &=& {dt_n\over a_n} {\cal G}_{i,k,n;j,k'}, \hbox{ for}\ j\not= i, 
 \label{secondderivatives} \\
 S_{i,k,n;i,k',n}&=& (F^+_n + F^-_n)\delta_{k,k'} +  
{dt_n\over a_n}\left[ {\cal G}_{i,k,n;i,k'}
+{1\over 2}\Omega H_o^2\delta_{k,k'}
 \right].\nonumber
\eeqa
The derivatives of the acceleration for $i\not= j$ are
\beq
{\cal G}_{i,k,n;j,k'} = Gm_j\left(  {\delta_{k,k'}\over |x_{i,n} - x_{j,n}|^3} -
3{(x_{j,k,n}-x_{i,k,n})(x_{j,k',n}-x_{i,k',n}) \over 
|x_{i,n} - x_{j,n}|^5 }\right),
\eeq
for the inverse square law, and, for the isothermal sphere model,
\beq
{\cal G}_{i,k,n;j,k'} = Gm_j\left(  {\delta_{k,k'}\over x_o|x_{i,n} - x_{j,n}|^2} -
2{(x_{j,k,n}-x_{i,k,n})(x_{j,k',n}-x_{i,k',n}) \over 
x_o|x_{i,n} - x_{j,n}|^4}\right).
\eeq
For $i=j$ the derivatives are
\beq
{\cal G}_{i,k,n;i,k'} =  -\sum_{j\not= i} {\cal G}_{i,k,n;j,k'}.
\eeq

Equation~(\ref{eq:shifts}) with the nonzero terms eliminated is
\beq
S_{i,k,n} + S_{i,k,n;i,k, n+1}\delta x_{i,k,n+1}+ 
 \sum_{j,k'}S_{i,k,n;j,k',n}\delta x_{j,k',n}
+ S_{i,k,n;i,k,n-1}\delta x_{i,k,n-1}= 0.  \label{eq:eofm2}
\eeq
Setting $n+1\rightarrow n$ and rearranging gives
\beq
\delta x_{i,k,n} = - {S_{i,k,n-1}+ \sum_{j,k'} S_{i,k,n-1;j,k',n-1}\delta x_{j,k',n-1} 
+ S_{i,k,n-1;i,k,n-2}\,\delta x_{i,k,n-2}
\over S_{i,k,n-1;i,k,n} }, \label{eq:42}
\eeq
which iterates to
\beq
\delta x_{i,k,n} = A_{i,k,n} + \sum_{j',k\pp} B_{i,k,n;j',k\pp}\delta x_{j',k\pp,1}.\label{eq:43}
\eeq
At $n=1$,
\beq
 A_{i,k,1} = 0,\qquad B_{i,k,1;j',k\pp} = \delta_{i,j'}\delta_{k,k\pp}.
 \label{eq:n=1}
\eeq
At  $n=2$, equation (\ref{eq:42}) is
\beq
\delta x_{i,k,2} = - \bigg[S_{i,k,1}+ \sum_{j',k\pp} S_{i,k,1;j',k\pp,1}\delta x_{j',k\pp,1} \bigg]/
 S_{i,k,1;i,k,2} , \label{eq:45}
\eeq
so 
\beq
A_{i,k,2} = - S_{i,k,1}/S_{i,k,1;i,k,2},
\qquad B_{i,k,2;j,k\pp} = - S_{i,k,1;j,k\pp,1}/S_{i,k,1;i,k,2}.
\label{eq:46}
\eeq
At $n\geq 3$ the form (\ref{eq:43}) in equation (\ref{eq:42})  gives
\beqa
\delta x_{i,k,n} = &-& \bigg[S_{i,k,n-1} + 
\sum_{j,k'} S_{i,k,n-1;j,k',n-1}
\bigg( A_{j,k',n-1} + \sum_{j',k\pp} B_{j,k',n-1;j',k\pp}\delta x_{j',k\pp,1}\bigg) \nonumber \\
&+& S_{i,k,n-1;i,k,n-2}\bigg(A_{i,k,n-2} + \sum_{j',k\pp} B_{i,k,n-2;j',k\pp}\,\delta x_{j',k\pp,1}\bigg)\bigg]
/ S_{i,k,n-1;i,k,n} .\label{eq:47}
\eeqa
Thus at $3\leq n\leq n_x$ the coefficients are
\beqa
&&A_{i,k,n} = - {S_{i,k,n-1} + \sum_{j,k'} S_{i,k,n-1;j,k',n-1}A_{j,k',n-1}
+ S_{i,k,n-1;i,k,n-2}A_{i,k,n-2}\over S_{i,k,n-1;i,k,n}},\nonumber \\
&& B_{i,k,n;j',k\pp} = - {\sum_{j,k'} S_{i,k,n-1;j,k',n-1}B_{j,k',n-1;j',k\pp}
+ S_{i,k,n-1;i,k,n-2}B_{i,k,n-2;j',k\pp}\over S_{i,k,n-1;i,k,n}}. \label{eq:48}
\eeqa
This gives the coefficients $A_{i,k,n}$ and $B_{i,k,n;j',k\pp}$ by iteration starting from equations~(\ref{eq:n=1}) and~(\ref{eq:46}).

Equation (\ref{eq:eofm2}) at $n=n_x$, with $\delta x_{i,k,n_x+1}= 0$ because present  positions are fixed, is
\beqa
0 &=& S_{i,k,n_x}  +  \sum_{j,k'}S_{i,k,n_x;j,k',n_x}\delta x_{j,k',n_x}
+ S_{i,k,n_x;i,k,n_x-1}\delta x_{i,k,n_x-1} \nonumber\\
&=&  S_{i,k,n_x}  +  \sum_{j,k'}S_{i,k,n_x;j,k',n_x}
\bigg[ A_{j,k',n_x} + \sum_{j',k\pp} B_{j,k',n_x;,j',k\pp}\delta x_{j',k\pp,1}\bigg] \\
&& \qquad\ + S_{i,k,n_x;i,k,n_x-1}\bigg[
A_{i,k,n_x-1} + \sum_{j',k\pp} B_{i,k,n_x-1;j',k\pp}\delta x_{j',k\pp,1}\bigg], \nonumber\label{eq:}
\eeqa
or
\beqa
&&0 = T_{i,k} + \sum_{j',k\pp}T_{i,k;j',k\pp}\delta x_{j',k\pp,1}, \nonumber\\
&& T_{i,k} = S_{i,k,n_x}  +  \sum_{j,k'}S_{i,k,n_x;j,k',n_x}A_{j,k',n_x} 
+ S_{i,k,n_x;i,k,n_x-1}A_{i,k,n_x-1},  \\
&& T_{i,k;j',k\pp} = \sum_{j,k'}S_{i,k,n_x;j,k',n_x}B_{j,k',n_x;j',k\pp}
+ S_{i,k,n_x;i,k,n_x-1} B_{i,k,n_x-1;j',k\pp}.  \nonumber
\eeqa
Inversion of the matrix $T_{i,k;j',k\pp}$ fixes the $\delta x_{i,k,1}$,  and equations (\ref{eq:43}) and (\ref{eq:48}) give the rest of the $\delta x_{i,k,n}$.

\end{document}